\documentclass[twocolumn,pre,showpacs]{revtex4-1}
\usepackage{graphicx}
\usepackage{amsmath}
\usepackage{amsfonts}
\usepackage{mathtools}
\usepackage[utf8]{inputenc}
\usepackage{subfigure}
\usepackage{psfrag}

\begin{document}
\title{Thermodynamic Uncertainty Relation for Biomolecular Processes}

\author{Andre C. Barato and Udo Seifert}
\affiliation{ II. Institut f\"ur Theoretische Physik, Universit\"at Stuttgart, 70550 Stuttgart, Germany}

\parskip 1mm
\def\F{\mathcal{A}} 
\def\Q{\mathcal{Q}}

\begin{abstract}
Biomolecular systems like molecular motors or pumps, transcription and translation machinery, and 
other enzymatic reactions can be described as Markov processes on a suitable network. We show 
quite generally that in a steady state the dispersion of observables like the number of consumed/produced molecules or
the number of steps of a motor is constrained by the thermodynamic cost of generating it. An uncertainty  $\epsilon$ 
requires at least a cost of $2k_BT/\epsilon^2$ independent of the time required to generate the output.

\end{abstract}

\maketitle


Biomolecular processes are generally out of equilibrium and dissipative, with the associated free energy consumption coming 
most commonly from adenosine triphosphate (ATP) hydrolysis. The role of energy dissipation in a variety of processes
related to biological information processing has received much attention recently \cite{qian05c,qian07,tu08a,lan12,meht12,muru12,bara13b,skog13,gove13,beck13,sart13,muru14,lang14,bara14a},
to give just one class of examples for which one tries to uncover fundamental limits involving energy dissipation in biomolecular systems.
    
Chemical reactions catalyzed by enzymes are of central importance for many cellular processes. Prominent examples are molecular
motors \cite{kolo07,astu10,toya10,vond11,zimm12}, which convert chemical free energy from ATP into mechanical work.
In this case an observable of interest is the number of steps the motor made. Another commonly analyzed  
output in enzymatic kinetics is the number of product molecules generated by an enzymatic reaction, 
for which the Michaelis–Menten scheme provides a paradigmatic case \cite{qian07}.

Quite generally, chemical reactions are well described by stochastic processes. An observable, like the rate of consumed substrate molecules or the number of 
steps of a motor on a track, is a random variable subjected to thermal fluctuations. Single molecule experiments \cite{svob93,rito06,gree07,corn07,moff08} provide detailed quantitative data on such random quantities. 
Obtaining information about the underlying chemical reaction scheme through the measurement of fluctuations constitutes a field called statistical kinetics \cite{schn95,shae05,moff10,moff14}. A central result in this field 
is the fact that the Fano factor quantifying fluctuations provides a lower bound on the number of states involved in an enzymatic cycle \cite{kolo07,moff14}.
 
For a non-zero mean output, the chemical potential difference (or affinity) driving an enzymatic reaction must also be non-zero, leading to
a free energy cost. Is there a fundamental relation between the relative uncertainty associated with the observable quantifying the output and the free 
energy cost of sustaining the biomolecular process generating it? 

In this Letter, we show that such a general bound does indeed exist. Specifically, for any process running for a time $t$,
we show that the product of the average dissipated heat and the square of the relative uncertainty of a generic observable 
is independent of $t$ and bounded by $2 k_BT$. This uncertainty relation is valid for general networks
and can be proved within linear response theory. Beyond linear response theory, we show it analytically for unicyclic
networks and verify it numerically for several multicyclic networks. As an illustration of a specific consequence of our results, we obtain a 
new bound on the Fano factor for unicyclic networks which depends on the affinity driving the process.

The observables we consider here arise from counting the number 
of transitions between states, which is different from a random variable counting the fraction of time spent 
in a state. For this latter quantity, the relative uncertainty is finite even in equilibrium, as is the case of a cell estimating the concentration of an external 
environment by counting the fraction of time receptors are bound \cite{bial05,endr09}. The role of dissipation for this problem has been recently studied in \cite{meht12,lang14}.


Our main result can be introduced using the arguably simplest example of
a nonequilibrium chemical reaction catalyzed by an enzyme, which is a biased
random walk where a single step is interpreted as the completion of an enzymatic cycle.
Steps to the right happen with a rate $k^+$, those to the left with a rate $k^-$.  
After a time $t$, on average, $\langle X\rangle=(k^+-k^-)t$ steps have occurred, with the number of steps  $X$ 
corresponding to the observable of interest. Specific realizations of this random
process show a variance $\langle (X-\langle X\rangle)^2\rangle= (k^+ + k^-)t$ \cite{vankampen}. The squared
relative uncertainty of the observable is
\begin{equation}
\epsilon^2\equiv  (\langle X^2\rangle-\langle X\rangle^2)/\langle X\rangle^2= (k^++k^-)/[(k^+-k^-)^2t].
\label{1defeps}
\end{equation}
Assuming an external environment of fixed temperature $T$, the transition rates are given by the local detailed balance relation
\begin{equation}
k^+/k^-=\exp(\F/k_BT),
\label{DB}
\end{equation}
where $\F$ is the affinity driving the process and $k_B$ Boltzmann's constant. The thermodynamic cost of generating this output is given by the entropy production rate,
which reads \cite{seif12}
\begin{equation}
\sigma=(k^+-k^-) \F/T,
\label{1defent}
\end{equation}
leading to a total dissipation after time $t$ of $T\sigma t$. In equilibrium, i.e., for $k^+=k^-$, there is no dissipation and the uncertainty $\epsilon$ diverges.
 
The trade-off between precision and dissipation is captured by
the crucial product $\Q$ of total dissipation and the square of the relative uncertainty,
\begin{equation}
\Q\equiv T\sigma t \epsilon^2 = \F \coth[\F/(2k_B T)] \geq 2 k_B T,
\label{defR}
\end{equation}
where we used Eqs. (\ref{1defeps}), (\ref{DB}), and (\ref{1defent}). This thermodynamic uncertainty relation shows that a more precise
output requires a higher thermodynamic cost independent of the time
used to produce the output. Reaching an uncertainty of, e.g., one 
percent requires at least $20000 k_BT$ of free energy. Since $\Q$ is an increasing 
function of the affinity $\F$, the minimal cost for a given 
uncertainty is achieved close to equilibrium, i.e., for $\F \to 0$. In this limit, however, the 
time $t$ required for producing a substantial output $\langle X\rangle$ diverges.

In the following, we show that this uncertainty relation, namely, that the dissipation of a process that leads to
an uncertainty $\epsilon$ must be at least $2k_BT/\epsilon^2$, is quite general, holding true for arbitrary networks of states.
First, we prove $\Q\ge 2k_BT$ for any network within linear response theory. For a unicyclic network
we show analytically that the bound also holds true beyond linear response. For multicyclic networks beyond
the linear response regime, we provide numerical evidence for this bound. From now on, to keep notation slim,  we set $k_B=T=1$,
which  renders entropy and energy dimensionless.


We consider a general Markov process with transition rate from state $i$ to $j$ denoted by $k_{ij}$. Thermodynamic consistency requires that if $k_{ij}\neq 0$
then $k_{ji}\neq 0$. Furthermore, we assume a finite number of states $N$ and denote the stationary probability of state $i$ by $P_i$.  

The observable of interest $X_\alpha$ represents some physical quantity that changes if certain transitions in the network of states occur. Specifically,
the generalized distance $d_{ij}^\alpha$ determines how much the variable $X_\alpha$ changes if the transition $i$ to $j$ happens. As an example, if $X_\alpha$ counts the number
of consumed ATP molecules, and if state $i$ represents a free enzyme and $j$ an enzyme with ATP bound to it, then $d_{ij}^\alpha=1$ and $d_{ji}^\alpha=-1$. This
generalized distance is always antisymmetric in $i$ and $j$.  

The affinity associated with the variable $X_\alpha$ is denoted $\F_\alpha$. For example, if $X_\alpha$ is the number of consumed substrate molecules (like ATP) in a chemical 
reaction, then $\F_\alpha$ is the chemical potential difference driving this reaction. The transition rates fulfill the generalized detailed balance relation \cite{seif12}
\begin{equation}
\ln(k_{ij}/k_{ji})=\sum_\beta d_{ij}^\beta \F_\beta+E_i-E_j,
\end{equation}
where this sum is over all affinities $\F_\beta$, including the case $\beta=\alpha$, and $E_i$ is the equilibrium free energy of state $i$.

In the stationary state, the velocity (or probability current) and diffusion constant associated with $X_\alpha$ are defined as
\begin{equation}
J_\alpha\equiv \langle X_\alpha\rangle/t
\end{equation}
and
\begin{equation}
D_\alpha\equiv [\langle X_\alpha^2\rangle-\langle X_\alpha\rangle^2]/(2t),
\end{equation}
respectively. The squared relative uncertainty then reads  
\begin{equation}
\epsilon_\alpha^2\equiv [\langle X_\alpha^2\rangle-\langle X_\alpha\rangle^2]/\langle X_\alpha\rangle^2= 2D_\alpha/(J_\alpha^2 t).
\label{eqeps}
\end{equation}
While the probability current has a simple form in terms of the stationary probability distribution, namely, $J_\alpha=\sum_{ij} d_{ij}^\alpha (P_i k_{ij}-P_j k _{ji})$, a general formula for 
the diffusion constant is more involved and will be discussed below. The entropy production rate is \cite{seif12}
\begin{equation}
\sigma= \sum_\beta J_\beta\F_\beta,
\label{genent}
\end{equation}
where $\beta$ runs over all affinities. For example, for a molecular motor this sum has two terms: one affinity is the chemical potential difference driving the motor 
with the rate of ATP consumption as the associated current, the other affinity is the mechanical force and the respective current is the velocity of the motor. 
The (dimensionless) product (\ref{defR}) for a general network is then defined as
\begin{equation}
\Q_\alpha\equiv \sigma t \epsilon_\alpha^2=  2\sigma D_\alpha/J_\alpha^2.
\label{ratio}
\end{equation}

Within linear response theory \cite{seif12}, which is valid close to equilibrium where the affinities $\F_\beta$ are small,
a current can be expressed by the affinities as 
\begin{equation}
J_\beta= \sum_\gamma L_{\beta\gamma}\F_\gamma,
\label{currlin}
\end{equation}
where the Onsager coefficients are defined as
\begin{equation}
L_{\gamma \beta}\equiv \left.\partial_{\F_\gamma} J_\beta\right|_{\F=0}=L_{\beta\gamma}.
\end{equation}
From Eqs. (\ref{genent}) and (\ref{currlin}) the entropy production 
within linear response reads $\sigma=\sum_{\beta,\gamma}L_{\beta\gamma}\F_\gamma\F_\beta$. Moreover, the diffusion constant is given by the Einstein relation $D_\alpha= L_{\alpha\alpha}$ \cite{andr07b},
which from Eq. (\ref{eqeps}) leads to $\epsilon_\alpha^{-1}$ being linear in the affinities. Hence, Eq. (\ref{ratio}) becomes
\begin{equation}
\Q_\alpha= 2\frac{\sum_{\beta,\gamma}L_{\alpha\alpha}L_{\beta\gamma}\F_\beta\F_\gamma}{\sum_{\beta,\gamma}L_{\alpha\beta}L_{\alpha\gamma}\F_\beta\F_\gamma}=
2\left(1+\frac{\sum_{\beta,\gamma\neq\alpha}G_{\beta\gamma}\F_\beta\F_\gamma}{(J_\alpha)^2}\right),
\label{eqrelin}
\end{equation}
where $G_{\beta\gamma}\equiv(L_{\alpha\alpha}L_{\beta\gamma}-L_{\alpha\beta}L_{\alpha\gamma})$. Using the fact that the Onsager matrix $L$ is positive semi-definite
it is possible to prove that $G$ is also a positive semi-definite matrix \cite{supp}. Hence, we have established $\Q_\alpha\ge 2$ within linear response theory. Note that equality is 
reached in the case of only one non-zero affinity, i.e., $\F_\beta=0$ for $\beta\neq \alpha$.

In the calculations that follow we use elegant expressions obtained by Koza \cite{koza99,koza02} 
for velocity and diffusion coefficient, which are valid for a general network of states. 
For these expressions we need a modified generator associated with $X_\alpha$, which is a $N$-dimensional square matrix with elements \cite{lebo99}
\begin{equation}	
[\mathcal{L}^\alpha(z)]_{ij}=\left\{\begin{array}{ll} 
 k_{ij}\exp(z d_{ij}^\alpha) & \quad \textrm{if } i\neq j\\
 -\sum_jk_{ij} & \quad \textrm{if } i=j
\end{array}\right.\,.
\label{modgenerator}
\end{equation}
A set of coefficients $C_n(z)$  is defined  through the characteristic polynomial of this matrix as
\begin{equation}
\det\left(yI-\mathcal{L}^\alpha(z)\right)\equiv \sum_{n=0}^{N}C_n(z)y^n,
\label{eqdet}
\end{equation}
where $I$ represents the identity matrix. Using these coefficients, which are functions of the transition rates, the current and diffusion coefficient can  be written as \cite{koza99} 
\begin{equation}
J_\alpha= -C_0'/C_1,
\label{eqJC}
\end{equation}
and
\begin{equation}
D_\alpha= -(C_0''+2C_1'J_\alpha+2C_2 J_\alpha^2)/(2C_1),
\label{eqDC}
\end{equation}
where $C_n\equiv C_n(0)$ and the primes denote derivatives with respect to $z$ taken at $z=0$. A full derivation for these expressions is given in \cite{supp}.


We first consider an arbitrary uni-cyclic model with $N$ states \cite{derr83}. The transition rate from state $n$ to state $n+1$ ($n-1$) is denoted $k_n^+$ $(k_n^-)$, where $n=0,1,\ldots,N-1$.
The output $X$ counts the number of completed cycles. It is sufficient to count the number of transitions through one of the links in the cycle, which we choose to be the link
between states $0$ and $1$. The generalized distance associated with $X$ is then $d_{01}=-d_{10}=1$ and $d_{ij}=0$ for $ij\neq 01$. The cycle affinity is 
\begin{equation}
\F= \ln(\Gamma_+/\Gamma_-),
\label{cycleaff}
\end{equation} 
where $\Gamma^+\equiv \prod_{i=0}^{N-1}k_i^+$ and $\Gamma^-\equiv \prod_{i=0}^{N-1}k_i^-$.
An example of such unicyclic machine with $N=3$ is an enzyme $E$ that consumes ATP according to the scheme
\begin{equation}
 E+ATP\xrightleftharpoons[k_1^-]{k_0^+} ET \xrightleftharpoons[k_2^-]{k_1^+} ED+P_i\xrightleftharpoons[k_0^-]{k_2^+} E+ADP+P_i,
\label{eqreaction} 
\end{equation}
where ADP stands for adenosine diphosphate, $\textrm{P}_i$ for phosphate, $ET$ ($ED$) represents the enzyme with an ATP (ADP) bound to it.
In this case the variable $X$ is the number of consumed ATP molecules and the affinity is given
by the chemical potential difference $\F=\mu_{ATP}-\mu_{ADP}-\mu_{P}$. 

We can show that for a given number of states $N$ and affinity $\F$ the product $\Q$ reaches its minimal value for uniform rates, i.e., $k_i^+=(\Gamma^+)^{1/N}$ and $k_i^-=(\Gamma^-)^{1/N}$ independent of $i$,
leading to the bound \cite{supp} 
\begin{equation}
\Q\ge (\F/N)\coth[\F/(2N)]\ge 2,
\label{strongerbound}
\end{equation}
which is in agreement with Eq. (\ref{defR}) that corresponds to $N=1$. The bound (\ref{strongerbound}) gives the
minimal dissipation required to realize an uncertainty $\epsilon$ for given affinity $\F$ and number of states $N$. 
This bound is an increasing function of $\F$, hence $\Q$ is minimal for $\F\to0$ where $\Q\to 2$. 
A related quantity, defined as the ratio of a ``barometric'' force and an ``Einstein'' force, has been considered in \cite{fish99}, where a bound similar to $\Q\ge 2$ has been shown to hold 
for the case $N=2$ within a calculation keeping terms up to second order in the affinity $\F$. 

We now turn to a specific example showing how this new constraint involving fluctuations and 
energetic cost can be turned into a diagnostic tool to unveil a structural property of an 
enzymatic cycle. A quantity closely related to the relative uncertainty is the Fano factor
\begin{equation}
F\equiv  [\langle X^2\rangle-\langle X\rangle^2]/\langle X\rangle= 2D/J,
\end{equation}
which gives a measure of the dispersion associated with $X$, where $X$ counts the output of an enzymatic cycle. 
For unicyclic networks, this Fano factor is known to be bounded from below by $1/N$ \cite{kolo07,moff14}. Measurements of the Fano factor
can then be used to obtain a bound on the number of states of an underlying enzymatic cycle \cite{moff14}.
Our new bound (\ref{strongerbound}) implies
\begin{equation}
F=\Q/\F\ge  (1/N)\coth[\F/(2N)].
\label{strongerboundfano}
\end{equation}
For a diverging affinity, which is the case in chemical reaction schemes where at least one backward transition
rate is assumed to be zero, this bound becomes the known one $F\ge 1/N$.  For experiments where substrate and product concentrations are kept fixed 
and, consequently,  the value of the affinity  is known, as for example in \cite{toya11}, our stronger bound in Eq. (\ref{strongerboundfano}) 
constrains even further the number of states in such an enzymatic cycle.

\begin{figure}
\includegraphics[width=52mm]{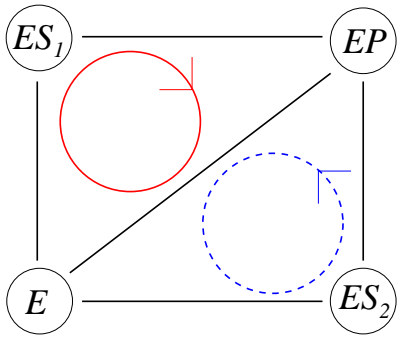}
\vspace{-2mm}
\caption{Multicyclic network of states for an enzyme $E$ that can bind substrates $S_1$ and $S_2$, and produces $P$.
The solid (dashed) cycle represents the first (second) cycle in Eq. (\ref{eqreaction2}). 
}
\label{fig1} 
\end{figure}

Let us turn again to multicyclic networks. Within the linear response regime, we have established above (after Eq. (\ref{eqrelin})) that $\Q_\alpha$ reaches 
the bound $2$ for the case where all affinities but $\F_\alpha$ are zero. For unicyclic networks, which is the paradigmatic case for a system with one affinity,
we have proved that the bound holds arbitrarily far from equilibrium, being reached only in the linear response regime. In order to provide full evidence that our
main result $\Q_\alpha\ge2$ is indeed universal, we now analyze multicyclic networks beyond linear response. In this case, we have to take specific systems. As a first example we consider a model with an enzyme $E$ 
that can consume two different substrates $S_1$ and $S_2$ and generates 
product $P$, see Fig. \ref{fig1}. Two enzymatic cycles of this model are 
\begin{eqnarray}
E+S_1\xrightleftharpoons[k_{21}]{k_{12}}ES_1\xrightleftharpoons[k_{42}]{k_{24}}EP\xrightleftharpoons[k_{14}]{k_{41}} E+P\nonumber\\
E+S_2\xrightleftharpoons[k_{31}]{k_{13}}ES_2\xrightleftharpoons[k_{43}]{k_{34}}EP\xrightleftharpoons[k_{14}]{k_{41}} E+P,
\label{eqreaction2} 
\end{eqnarray}
where the enzyme states are identified as  $E\mathrel{\hat=} 1$, $ES_1\mathrel{\hat=} 2$, $ES_2\mathrel{\hat=} 3$ and $EP\mathrel{\hat=} 4$.
The affinity of the cycle involving substrate $S_1$ ($S_2$) is given by the chemical potential 
difference $\F_1= \mu_1-\mu_P$ ($\F_2= \mu_2-\mu_P$). The relations between these affinities and the transition rates are $\F_1= \ln[k_{12}k_{24}k_{41}/(k_{21}k_{42}k_{14})]$ and
$\F_2= \ln[k_{13}k_{34}k_{41}/(k_{31}k_{43}k_{14})]$. There is also a third cycle $1\to2\to4\to3\to1$, in which an $S_1$ is consumed and
an $S_2$ produced. Its affinity is not independent but rather given by $\F_1-\F_2$. 

The average rate at which $S_1$ molecules are consumed is given by the current $J_1\equiv P_1k_{12}-P_2k_{21}$, while the rate of $S_2$ consumption is $J_2\equiv P_1k_{13}-P_3k_{31}$. 
The entropy production can be conveniently written as a sum over terms which are the product of 
a cycle affinity and a current \cite{seif12}. For the present model it becomes
\begin{equation}
\sigma= \F_1 J_1+\F_2 J_2. 
\label{ent4}
\end{equation}

As an observable of interest we choose the number of consumed $S_1$ molecules, for which we can calculate the associated product $\Q_1$ with    
formulas (\ref{ratio}), (\ref{eqJC}), (\ref{eqDC}), and (\ref{ent4}). The resulting function of the ten transition rates is
too cumbersome to show. Both by minimizing this function numerically and by evaluating it for randomly chosen transition rates
we find that the uncertainty relation $\Q_1\ge 2$ is respected. In order to verify whether this result is particular to
this network topology we have analyzed six other multicyclic networks, which do not share any particular symmetry \cite{supp}. 
For all these networks, our uncertainty relation $\Q_\alpha\ge 2$ is fulfilled. In all cases numerical minimization of $\Q_\alpha$ 
leads to a minimum compatible with $2$ that is reached in the linear response regime. Based on these results, we conjecture 
that $\Q_\alpha\ge 2$ also for general multicyclic networks beyond linear response.


In conclusion, for nonequilibrium stationary states we have conjectured the 
fundamental limit $2 k_BT/\epsilon^2$ on the
minimal dissipation required to generate an output with small relative
uncertainty $\epsilon$. This bound can be saturated close to equilibrium with
only one independent affinity driving the process. This uncertainty
relation provides a universal link between the ``precision'' of a molecular
machine and the cost of maintaining it. As one specific application we have
shown how a bound on the number of states involved in the enzymatic cycle
related to the Fano factor can be improved provided the affinity is known.
More broadly, one can expect similar signatures of our fundamental relation 
quantifying the minimal energetic cost for reaching small uncertainty, i.e.,
high precision, to show up in any biochemical or biophysical process 
at fixed temperature that can be described by a Markov network \cite{foot}. 
Exploring whether and how this balance between fluctuations and energetics
has guided the evolution of chemical reaction networks in living systems 
constitutes one intriguing perspective of our approach. Our fundamental 
relationship between minimal dissipated heat and uncertainty is based, first, 
on exact results in the linear response regime. Second, we have proved it 
for unicyclic networks arbitrarily far from equilibrium. Third, for multicyclic
networks far from equilibrium we have  numerical evidence for several different 
networks. We could not provide a formal proof for arbitrary networks, and we expect that 
the method used for unicyclic networks cannot be generalized to 
multicyclic networks, as it requires an expression for the diffusion coefficient in 
terms of the transition rates.    

On the technical level, investigating possible generalizations of the affinity 
dependent bound on the Fano factor in Eq. (\ref{strongerboundfano}) 
to multicyclic networks could lead to further new bounds in statistical kinetics.  
Likewise, it would be interesting to explore
whether one can derive bounds involving higher order cumulants.
Finally, we emphasize that an algebraic proof of the uncertainty relation
in the multicyclic case beyond the linear response regime looks like a serious 
challenge.




\begin{thebibliography}{10}
\bibitem{foot} Similarly, the generalization to machines involving coupling to two
heat baths should be straightforward.
\end{thebibliography}

\clearpage

\onecolumngrid

\section*{Supplemental material: Thermodynamic uncertainty relation for biomolecular processes}

\section{Proving that the matrix $G$ is positive semi-definite}

We prove that the matrix $G$ in Eq. (13) in the main text is positive semi-definite.
For a given $n$ by $n$ Onsager matrix $L$, where $n$ is the number of independent affinities, we define the $(n-1)$-dimensional matrix
$G$ through
\begin{equation}
G_{\beta\gamma}\equiv L_{11}L_{\beta \gamma}- L_{1\beta}L_{1 \gamma}
\end{equation}
where $\beta,\gamma= 2,\ldots,n$. For convenience and without loss of generality we have set $\alpha=1$ in comparison with the main text.
Since $L$ is positive semi-definite, $G$ is symmetric and its diagonal elements are positive, which are necessary conditions for a positive definite matrix.

From Sylvester's identity for bordered determinants \cite{matrixbook} it follows that 
\begin{equation}
\det(G)= L_{11}^{n-2}\det(L).
\label{Sylv}
\end{equation}
This relation guarantees that $\det(G)\ge 0$. A principal minor of $G$ is given by the determinant
of the matrix that is obtained by eliminating columns and lines from $G$. For example, a principal
minor of order $1$, denoted by $g_\beta^{(1)}$ is given by the determinant of the matrix that is obtained
by deleting column and line $\beta$ from matrix $G$, with $\beta=2,\ldots,n$. If we denote by  $l_\beta^{(1)}$
a principal minor of order one of the matrix $L$, then, similarly to Eq. (\ref{Sylv}), Sylvester's identity for bordered determinants gives
\begin{equation}
g_\beta^{(1)}= L_{11}^{n-3}l_\beta^{(1)}.
\label{Sylv2}
\end{equation}
Hence, we have that the principal minors of order one of the matrix $G$ are non-negative. Clearly, relations similar to (\ref{Sylv2})
are valid for any principal minor of order $k$, of which there are $(n-1)!/[k!(n-1-k)!]$. Therefore,
all principal minors of $G$ are non-negative which, from Sylvester's criterion, implies $G$ being positive semi-definite.

\section{Expressions for velocity and diffusion constant}

In this section we derive the expressions for velocity and diffusion obtained by Koza \cite{koza99}.
The modified generator $\mathcal{L}^\alpha(z)$ is a $N$ by $N$ matrix with elements
\begin{equation}	
[\mathcal{L}^\alpha(z)]_{ij}\equiv\left\{\begin{array}{ll} 
 k_{ij}\exp(z d_{ij}^\alpha) & \quad \textrm{if } i\neq j\\
 -\sum_jk_{ij} & \quad \textrm{if } i=j
\end{array}\right.\,.
\label{modgenA}
\end{equation}
It can be shown that the maximum eigenvalue of this generator $\lambda(z)$ gives the scale cumulant generating function related 
to the random variable $X_\alpha$ \cite{lebo99}. The current and diffusion constant can be obtained from $\lambda(z)$ with the formulas
\begin{equation}
J_\alpha= \lambda'
\label{Jcum}
\end{equation}
and
\begin{equation}
D_\alpha= \lambda''/2,
\label{dcum}
\end{equation}
where the primes denote derivatives taken at $z=0$. 
Furthermore, the characteristic polynomial associated with $\mathcal{L}^\alpha(z)$ reads
\begin{equation}
P(z,y)\equiv\det\left(yI-\mathcal{L}^\alpha(z)\right)= \sum_{n=0}^{N}C_n(z)y^n.
\label{Pzy}
\end{equation}
Since $\lambda(z)$ is a root of this characteristic polynomial, the following relation holds
\begin{equation}
\sum_{n=0}^{N}C_n(z)\lambda^n(z)=0.
\label{chac1}
\end{equation}
Moreover, for $z=0$ the matrix in Eq. (\ref{modgenA}) becomes an stochastic matrix leading to $\lambda(0)=0$. Taking the derivative with respect to $z$ and setting $z=0$ in Eq. (\ref{chac1}) 
leads to
\begin{equation}
\lambda'= -C_0'/C_1.
\label{firstdev}
\end{equation}
Taking a second derivative of Eq. (\ref{chac1}) with respect to $z$ and setting $z=0$ gives 
\begin{equation}
\lambda''= -[C_0''+2C_1'\lambda'+2C_2(\lambda')^2]/(C_1).
\label{seconddev}
\end{equation}
Using Eqs. (\ref{Jcum}), (\ref{dcum}), (\ref{firstdev}), and (\ref{seconddev}), we obtain the final formulas
\begin{equation}
J_\alpha= -C_0'/C_1,
\label{relaJ}
\end{equation}
and
\begin{equation}
D_\alpha= -(C_0''+2C_1'J_\alpha+2C_2 J_\alpha^2)/(2C_1),
\label{relaD}
\end{equation}
which are Eqs. (16) and (17) in the main part.

\section{Calculations for unicyclic network}
\label{appa}

For  a unicyclic network, the coefficient $C_0(z)$ in Eq. (\ref{Pzy}) is given by the determinant of the matrix (\ref{modgenA}), which 
reads $C_0(z)= -(\textrm{e}^z-1)\Gamma_-(\textrm{e}^\F-\textrm{e}^{-z})$.
Therefore, with Eqs. (\ref{relaJ}), (\ref{relaD}), and the entropy production for the unicyclic model $\sigma= J\F$, the quantity $\Q$ becomes 
\begin{equation}
\Q= \frac{2D \F}{J}= \frac{\textrm{e}^\F+1}{\textrm{e}^\F-1}\F-2\frac{C_2 \Gamma_-}{C_1^2} (\textrm{e}^\F-1)\F,
\label{Runi}
\end{equation}
where we used the fact that $C_1'=0$ for a unicyclic network \cite{koza99}.
We now show that $C_2 \Gamma_-/C_1^2$ reaches its maximum when $k_i^+=(\Gamma^+)^{1/N}$ and $k_i^-=(\Gamma^-)^{1/N}$ for $i=0,\ldots,N-1$.
It is convenient to write the rates in the form
\begin{equation}
k_i^+= \phi_i\textrm{e}^{\F\theta_i/2}\qquad\textrm{and}\qquad k_i^-= \phi_i\textrm{e}^{-\F\theta_i/2},
\end{equation}
where the constraint $\sum_{i=0}^{N-1}\theta_i=1$ fixes the affinity $\F$.   
We define the function
\begin{equation}
h(n,s)\equiv (\Phi/\phi_s)\textrm{e}^{\F\Theta(n,s)/2}
\end{equation}
where
\begin{equation}	
\Theta(n,s)\equiv\left\{\begin{array}{ll} 
 \sum_{i=n+1}^{s-1}\theta_i-\sum_{i=s+1}^{N-1}\theta_i-\sum_{i=0}^{n}\theta_i & \quad \textrm{if } s\ge n+1\\
 \sum_{i=n+1}^{N-1}\theta_i+\sum_{i=0}^{s-1}\theta_i-\sum_{i=s+1}^{n}\theta_i & \quad \textrm{if } s\le n,
\end{array}\right.\,
\label{teta}
\end{equation}
and 
\begin{equation}
\Phi\equiv \prod_{i=0}^{N-1}\phi_i.
\end{equation}
With this function the coefficients of the characteristic polynomial in Eq. (\ref{Pzy}) can be written as
\begin{equation}
C_1= \sum_{n=0}^{N-1}\sum_{s=0}^{N-1}h(n,s),
\label{C1exp}
\end{equation}
and
\begin{equation}
 C_2\Gamma_-=  \sum_{l=1}^{N-1}\sum_{n=0}^{N-l-1}\sum_{s=n+1}^{n+l}h(n,s)\bigg(\sum_{t=0}^{n}h(n+l,t)+\sum_{t=n+l+1}^{N-1}h(n+l,t)\bigg)
\label{C2exp}
\end{equation}
These two relations can be obtained from the formula \cite{koza99}
\begin{equation}
C_l= \sum_{a,b}\prod_{m,n=0}^{N-1}(k^+_m)^{a_m}(k^-_n)^{b_n}\psi_l(a,b)
\end{equation} 
where $l=1,2$, $a$ ($b$) denotes a vector with components $a_n\in \{0,1\}$ ($b_n\in \{0,1\}$) and $\psi_l(a,b)\in \{0,1\}$. The function
$\psi_l(a,b)$ is non-zero only if $\sum_{n=0}^{N-1}(a_n+ b_n)=N-l$ and if for all $n=0,1,\ldots,N-1$ the relations $a_n=b_n=1$ and $a_n=b_{n+1}=1$ are not fulfilled.
Moreover, a graphical representation of Eq. (\ref{C1exp}) can be found in \cite{koza02}. 

To find the maximum of $f(\{\theta_i\},\{\phi_i\})\equiv C_2\Gamma_-/C_1^2$  we consider the function
\begin{equation}
\Lambda(\{\theta_i\},\{\phi_i\},\lambda_0)\equiv f(\{\theta_i\},\{\phi_i\})+\lambda_0\left(1-\sum_{i=0}^{N-1}\theta_i\right),
\end{equation}
where $\lambda_0$ is the Lagrange multiplier. This function is maximized for $\{\theta_i^*\},\{\phi_i^*\}$, which are given by
the solution of the equations
\begin{equation}
\partial_{\phi_j}f(\{\theta_i^*\},\{\phi_i^*\})= 0
\label{deriphi}
\end{equation}
and
\begin{equation}
\partial_{\theta_j}f(\{\theta_i^*\},\{\phi_i^*\})= \lambda_0,
\label{derithe}
\end{equation}
for $j=0,1,\ldots,N-1$. Before taking derivatives of the function $f$ it is convenient to define
\begin{equation}
h_j(n_j,s_j)\equiv h(n,s),
\end{equation} 
where $n_j= (n-j+N)\mod N$, $s_j= (s-j+N)\mod N$, and $j=0,1,\ldots,N-1$. Due to the symmetry of the unicyclic network, we can rewrite
Eqs (\ref{C1exp}) and (\ref{C2exp}) in the forms
\begin{equation}
C_1= \sum_{n=0}^{N-1}\sum_{s=0}^{N-1}h_j(n,s),
\label{C1exp2}
\end{equation}
and
\begin{equation}
C_2\Gamma_-=  \sum_{l=1}^{N-1}\sum_{n=0}^{N-l-1}\sum_{s=n+1}^{n+l}h_j(n,s)\bigg(\sum_{t=0}^{n}h_j(n+l,t)+\sum_{t=n+l+1}^{N-1}h_j(n+l,t)\bigg),
\label{C2exp2}
\end{equation}
which are valid for all $j$. A first derivative of the function $f$ reads
\begin{equation}
\partial_x f= \frac{\partial_x(C_2\Gamma_-)}{C_1^2}-2\frac{(\partial_xC_1)C_2\Gamma_-}{C_1^3},
\label{gendev}
\end{equation}
where $x=\phi_j,\theta_j$. Taking derivatives of Eqs. (\ref{C1exp2}) and (\ref{C2exp2}) with respect to $\phi_j$, leads to
\begin{equation}
\partial_{\phi_j}C_1=\frac{1}{\phi_j}\left(C_1-\sum_{n=0}^{N-1}h_j(n,0)\right)
\end{equation}
and
\begin{equation}
\partial_{\phi_j}C_2\Gamma_-=\frac{1}{\phi_j}\left[2C_2\Gamma_--\sum_{l=1}^{N-1}\sum_{n=0}^{N-l-1}\sum_{s=n+1}^{n+l}h_j(n,s)h_j(n+l,0)\right].
\end{equation}
At the symmetric point $\theta_i^*=1/N$ and $\phi_i^*=\phi$ for all $i$, these derivatives become
\begin{equation}
\partial_{\phi_j}C_1=\frac{1}{\phi}\frac{N-1}{N}C_1
\end{equation}
and
\begin{equation}
\partial_{\phi_j}C_2\Gamma_-=\frac{2}{\phi}\frac{N-1}{N}C_2\Gamma_-.
\end{equation}
Using these two last relations in Eq. (\ref{gendev}) we obtain that $\partial_{\phi_j}f=0$ at $\theta_i^*=1/N$ and $\phi_i^*=\phi$. Taking derivatives of  Eqs. (\ref{C1exp2}) and (\ref{C2exp2}) with respect to $\theta_j$ leads to
\begin{equation}
\partial_{\theta_j}C_1= \frac{\F}{2}\sum_{n=0}^{N-1}\left(\sum_{s=1}^{n}h_j(n,s)-\sum_{s=n+1}^{N-1}h_j(n,s)\right)
\end{equation}
and
\begin{equation}
\partial_{\theta_j}C_2\Gamma_-= -\frac{\F}{2}\sum_{l=1}^{N-1}\sum_{n=0}^{N-l-1}\sum_{s=n+1}^{n+l}h_j(n,s)\left(2\sum_{t=n+l+1}^{N-1}h_j(n+l,t)+h_j(n+l,0)\right).
\end{equation}
From these two relations and Eq. (\ref{gendev}) we obtain that the derivative $\partial_{\theta_j}f$ is independent of $j$ for $\theta_i^*=1/N$ and $\phi_i^*=\phi$ for all $i$.
In this case the Lagrange multiplier in Eq. (\ref{derithe}) becomes
\begin{equation}
\lambda_0= \frac{A \F}{2(\textrm{e}^{\F/N}-1)^2(\textrm{e}^{\F}-1)^3 N^2},
\end{equation}
where 
\begin{eqnarray}
A= 2 \textrm{e}^{\F/N}  -\textrm{e}^{2(N+1)\F/N}(N-1)N-\textrm{e}^{2\F}N(N+1)-\textrm{e}^{2(N+1)\F/N}N(3N+1)+\nonumber\\
\textrm{e}^{\F}(N-3N^2)+\textrm{e}^{(2N+1)\F/N}(N^2+1)+\textrm{e}^{(N+1)\F/N}(6N^2-4). 
\end{eqnarray}
Hence, we have shown that  for symmetric rates, with  $\theta_i^*=1/N$ and $\phi_i^*=\phi$ for all $i$, Eqs. (\ref{deriphi}) and (\ref{derithe}) are satisfied. From the explicit calculations of
the derivatives they seem to be the unique solution, however, we are not able to provide a rigorous proof of this uniqueness. The maximal value of the function $f$ is
\begin{equation}
\frac{C_2 \Gamma_-}{C_1^2}= \frac{(\textrm{e}^{\F+\F/N}-1)(N-1)-(\textrm{e}^{\F}-\textrm{e}^{\F/N})(N+1)}{2N(\textrm{e}^{\F/N}-1)(\textrm{e}^{\F}-1)^2}.
\end{equation}
Inserting this expression in Eq. (\ref{Runi}) leads to the lower bound in Eq. (20) in the main text.

\section{Verifying the bound $\mathcal{Q}\ge 2$ for multicyclic networks}

\begin{figure}
\subfigure[]{\includegraphics[width=32mm]{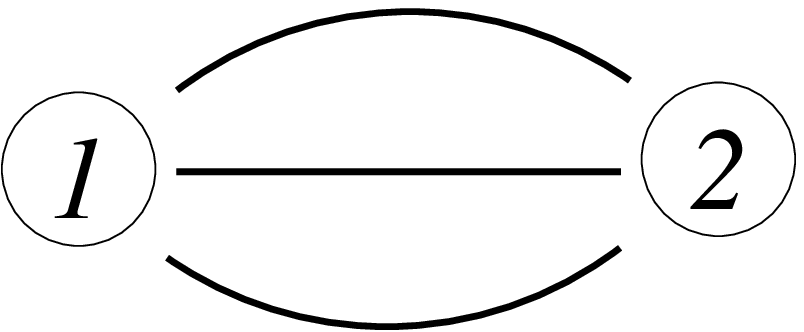}\label{figA1a}}\hfill
\subfigure[]{\includegraphics[width=52mm]{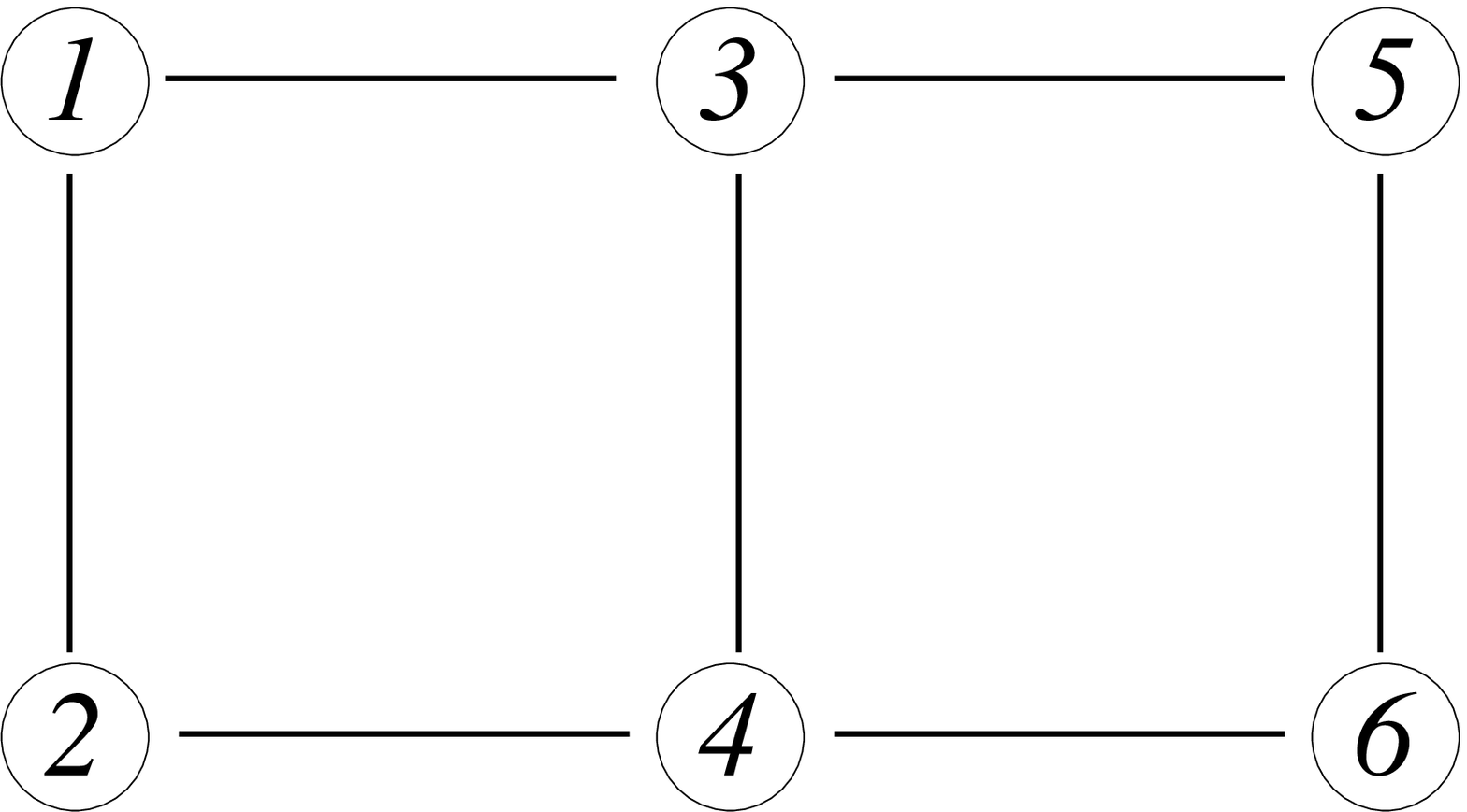}\label{figA1b}}\hfill
\subfigure[]{\includegraphics[width=52mm]{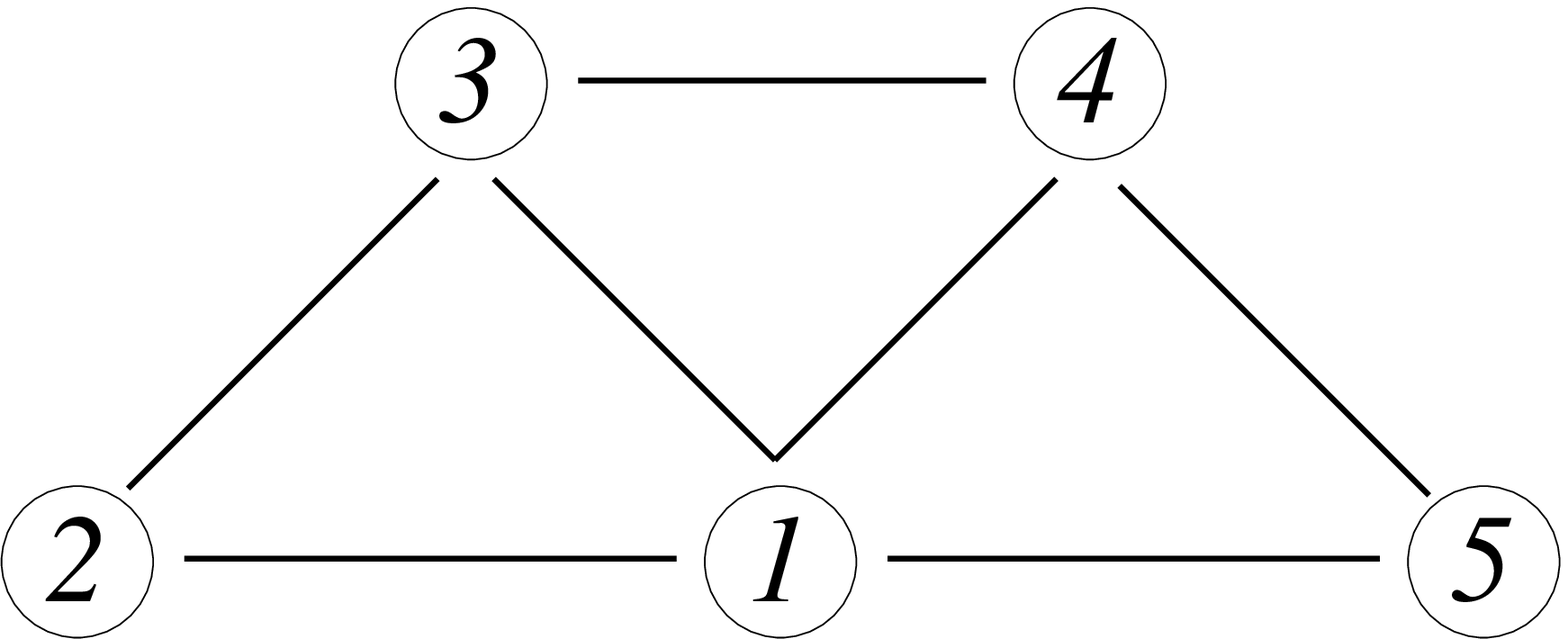}\label{figA1c}}
\subfigure[]{\includegraphics[width=52mm]{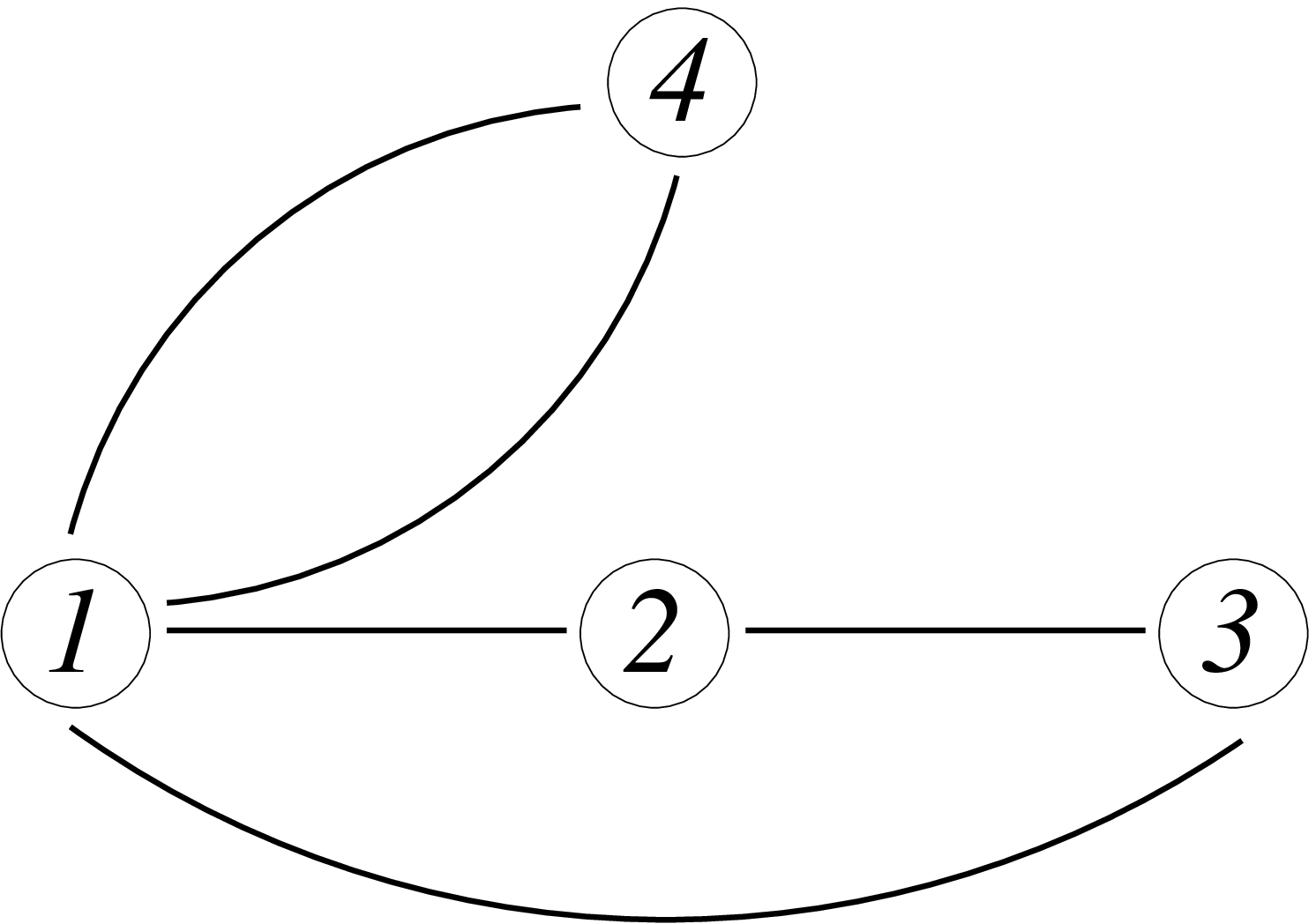}\label{figA1d}}\hfill
\subfigure[]{\includegraphics[width=32mm]{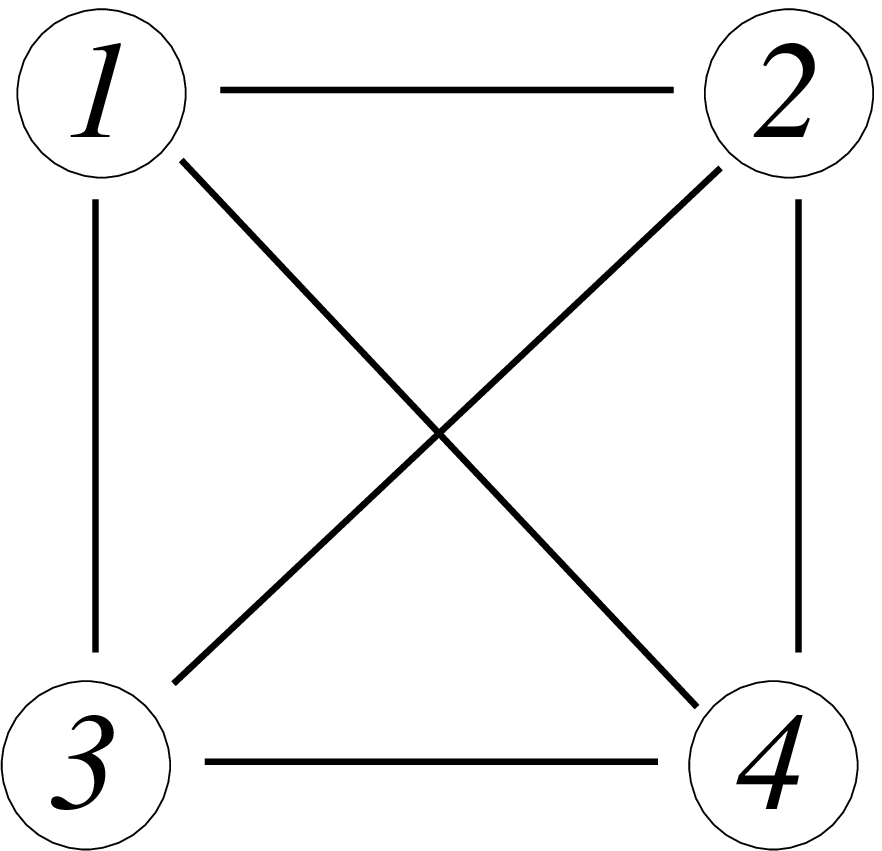}\label{figA1e}}\hfill
\subfigure[]{\includegraphics[width=52mm]{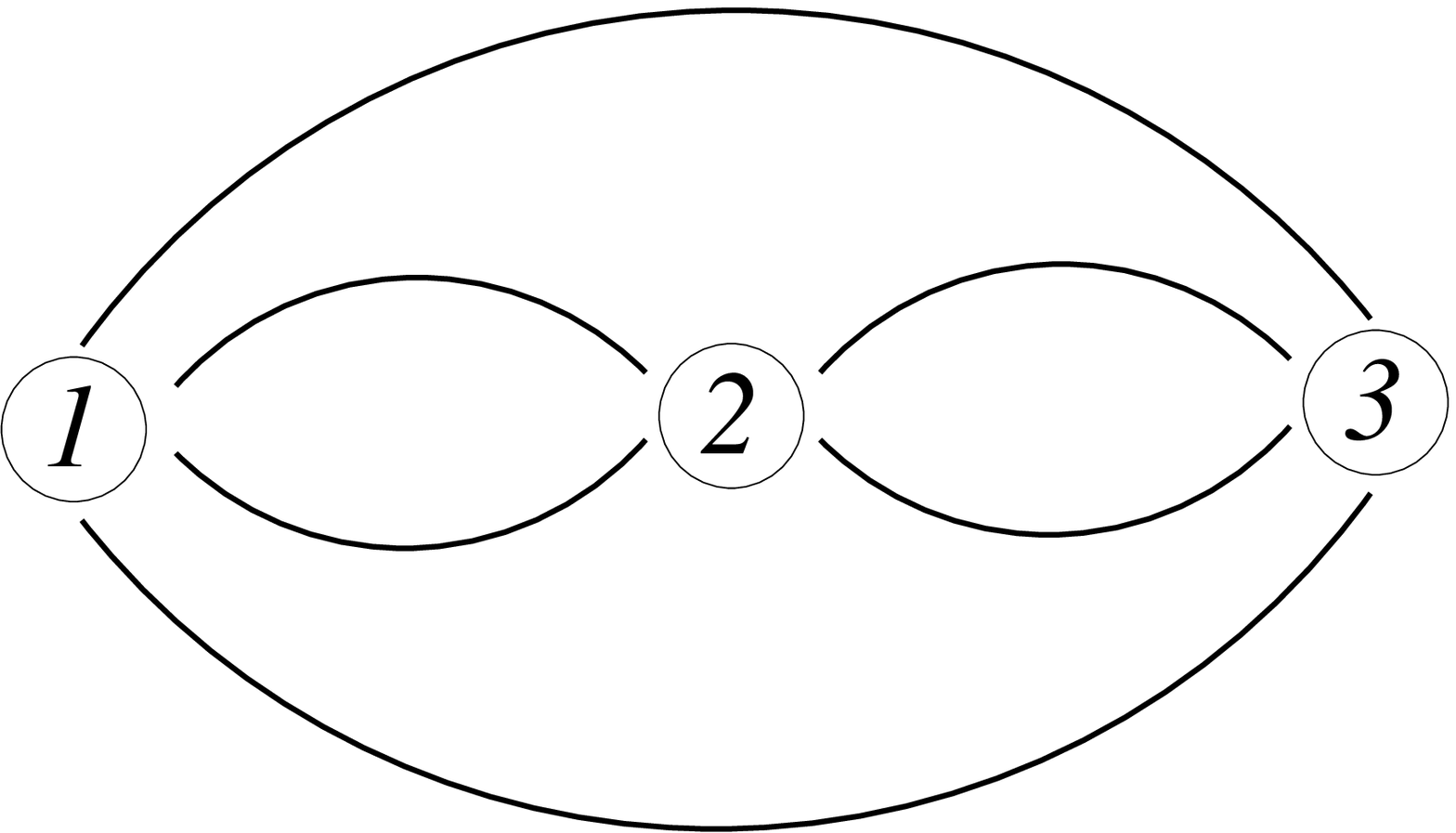}\label{figA1f}}
\vspace{-2mm}
\caption{Network of states for multicyclic models. (a) Two-state model with two fundamental cycles, (b) six-state model with two fundamental cycles,  (c) five-state model with three fundamental cycles, (d) 
four-state model with two fundamental cycles, (e) fully-connected four-state model, (f) three-state model with four fundamental cycles.}
\label{figA1} 
\end{figure}

We present numerical results supporting the uncertainty relation for six different models.
The simplest example of a multicyclic network is shown in Fig. \ref{figA1a}. The two states are denoted by $1$ and $2$
and the transition rates between then are $k_{12}^{\nu}$ and $k_{21}^{\nu}$, with $\nu=a,b,c$ for the three different links.
We take the output to be the number of transitions from $1$ to $2$ minus the number of transitions from $2$ to $1$ in link $a$.
With this choice the modified generator in Eq. (\ref{modgenA}) becomes
\begin{equation}
\left(
\begin{array}{cc}
-(k^a_{12}+k^b_{12}+k^{c}_{12}) & k^a_{12}\textrm{e}^z+k^b_{12}+k^{c}_{12} \\
k^a_{21}\textrm{e}^{-z}+k^b_{21}+k^{c}_{21}  & -(k^a_{21}+k^b_{21}+k^{c}_{21}) 
\end{array}
\right)  
\label{gen2}
\end{equation}

\begin{figure}
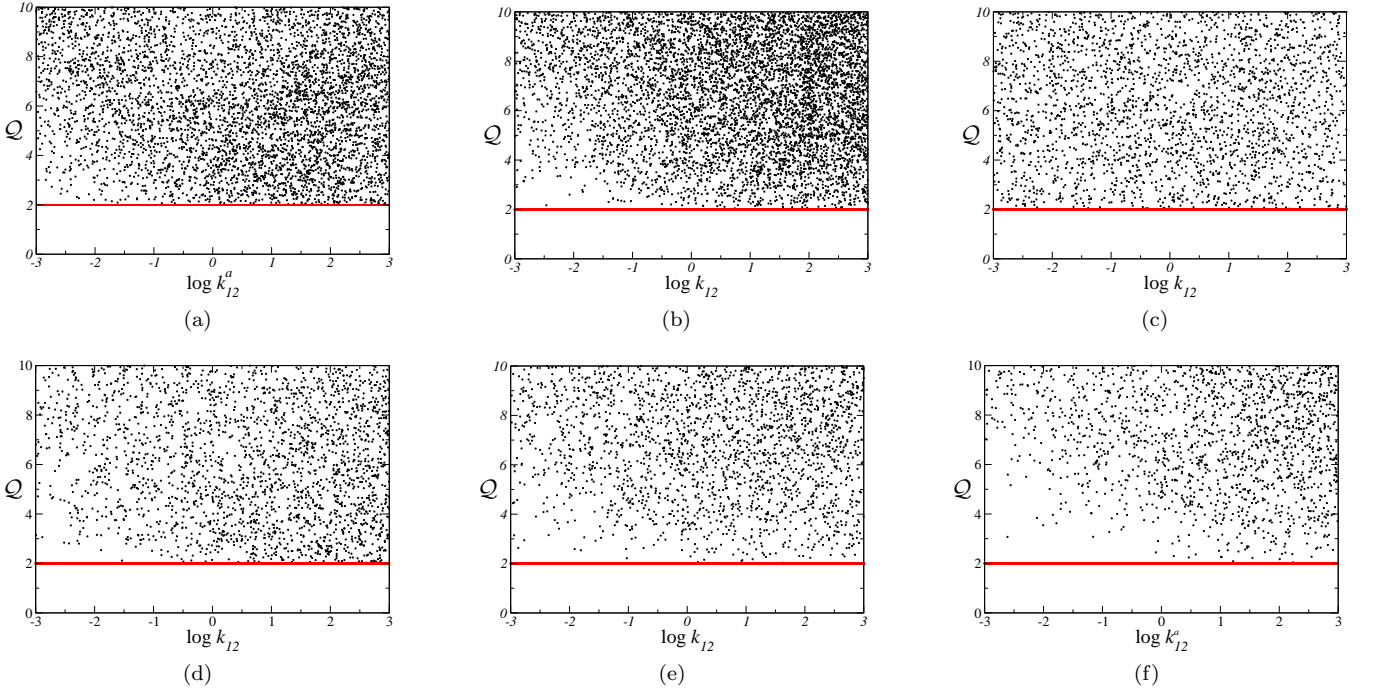

\subfigure[]{\psfrag{R}{$\mathcal{Q}$}\includegraphics[width=52mm]{./02statesbound2.eps}\label{figA2a}}\hfill
\subfigure[]{\psfrag{R}{$\mathcal{Q}$}\includegraphics[width=52mm]{./05statesbound2.eps}\label{figA2b}}\hfill
\subfigure[]{\psfrag{R}{$\mathcal{Q}$}\includegraphics[width=52mm]{./06states.eps}\label{figA2c}}\hfill
\subfigure[]{\psfrag{R}{$\mathcal{Q}$}\includegraphics[width=52mm]{./04satesimple2cy.eps}\label{figA2d}}\hfill
\subfigure[]{\psfrag{R}{$\mathcal{Q}$}\includegraphics[width=52mm]{./04fully.eps}\label{figA2e}}\hfill
\subfigure[]{\psfrag{R}{$\mathcal{Q}$}\includegraphics[width=52mm]{./03states4aff.eps}\label{figA2f}}
\vspace{-2mm}
\caption{Numerical check of the bound $\Q\ge 2$. (a) Network of Fig. \ref{figA1a} with output $X_{12}^a$, (b) network of Fig. \ref{figA1b} with output $X_{12}$, (c) network of Fig. \ref{figA1c} with output  $X_{23}$,
(d) network of Fig. \ref{figA1d} with output  $X_{12}$, (e) network of Fig. \ref{figA1e} with output  $X_{12}$, and (f) network of Fig. \ref{figA1f} with output  $X_{12}^a$. 
In all cases the transition rates $k_{ij}$ are randomly chosen by generating a random number $x$ between $-3$ and $3$ and then taking $k_{ij}=10^x$.}
\label{figA2} 
\end{figure}

The related current $J_a$ and diffusion coefficient $D_a$ are obtained from Eqs. (\ref{relaJ}) and (\ref{relaD}). They read
\begin{equation}
J_a= (K_1-K_2)/K_3
\label{eq2J}
\end{equation}
and
\begin{equation}
D_a= (K_1+K_2)/(2K_3)-(K_1-K_2)^2/(K_3)^3,
\label{eq2D}
\end{equation}
where $K_1\equiv k_{12}^a(k_{21}^b+k_{21}^c)$, $K_2\equiv k_{21}^a(k_{12}^b+k_{12}^c)$, and $K_3\equiv k_{12}^a+k_{12}^b+k_{12}^c+k_{21}^a+k_{21}^b+k_{21}^c$.
The entropy production for this model reads
\begin{equation}
\sigma= J_a\ln \frac{k_{12}^ak_{21}^b}{k_{21}^ak_{12}^b}+J_c\ln \frac{k_{12}^ck_{21}^b}{k_{21}^ck_{12}^b},
\label{eq2s}
\end{equation}
where $J_\nu= P_1k_{12}^\nu-P_2k_{21}^\nu$. With Eqs. (\ref{eq2J}), (\ref{eq2D}), and (\ref{eq2s}), we obtain $\Q_a= 2 D_a \sigma/ J_a$.
A numerical test of the bound $\Q_a\ge2$ is given in Fig. \ref{figA2a}, where we evaluate $\Q_a$ for randomly chosen transition rates.

Following the same procedure explained above for the two-state model we calculated $\mathcal{Q}$ for the other five network of states in Fig. \ref{figA1}. 
Numerical tests of the inequality  $\Q_\alpha\ge2$ for these networks are shown in Figs. \ref{figA2}, with the chosen observable $X_\alpha$ for each network indicated in the caption. 
Besides evaluating $\Q_\alpha$ at randomly chosen transition rates we have also minimized it numerically. In all cases the minimum value is compatible with $2$. Furthermore, the minimum is reached at the linear response regime with
the affinity $\F_{\alpha}$ small and the other affinities even smaller, compatible with $\F_{\beta}=0$ for $\beta\neq \alpha$. We point out that for the symmetric networks in Fig. \ref{figA1a}, 
Fig. \ref{figA1e}, and Fig. \ref{figA1f} all the currents are equivalent, i.e., all links connect the same number of nodes and all nodes have the same number of links.
For the other networks, we have also checked the bound for the following cases: outputs $X_{13}$ and $X_{34}$ 
for the network in Fig. \ref{figA1b}; outputs $X_{41}$, $X_{12}$, and $X_{34}$ for the network in Fig. \ref{figA1c}; outputs $X_{14}^a$ and $X_{23}$ for the network in Fig. \ref{figA1d}.

\end{document}